\documentclass[11pt,a4paper]{article}
\usepackage[utf8x]{inputenc}
\usepackage[T1]{fontenc}
\usepackage{gentium}
\usepackage{mathptmx} 
\usepackage[pdftex]{graphicx} 
\usepackage[pdftex,linkcolor=black,pdfborder={0 0 0}]{hyperref} 
\usepackage{calc} 
\usepackage{enumitem} 
\usepackage{multirow}
\usepackage{authblk}
\usepackage[a4paper, lmargin=0.12\paperwidth, rmargin=0.1666\paperwidth, tmargin=0.1111\paperheight, bmargin=0.1111\paperheight]{geometry} 
\usepackage{parskip}
\usepackage[all]{nowidow} 
\usepackage[protrusion=true,expansion=true]{microtype} 
\usepackage{units}
\usepackage{verbatim}
\usepackage{amsmath,amssymb}

\begin{document}
\title{Radial Density Profile and Stability of Capillary Discharge Plasma Waveguides of Lengths up to 40 Centimeters}
\author[1]{M.~Turner}
\author[1]{A.~J.~Gonsalves}
\author[1]{S.~S.~Bulvanov}
\author[1]{C.~Benedetti}
\author[2]{N.~A.~Bobrova}
\author[2]{V.~A.~Gasilov}
\author[2,3]{P.~V.~Sasorov}
\author[3]{G.~Korn}
\author[1]{K.~Nakamura}
\author[1]{J.~van Tilborg}
\author[1]{C.~G.~Geddes}
\author[1]{C.~B.~Schroeder}
\author[1]{E.~Esarey}

\affil[1]{Lawrence Berkeley National Laboratory, Berkeley, CA, USA}
\affil[2]{Keldysh Institute of Applied Mathematics RAS, Moscow, Russia}
\affil[3]{ELI Beamlines, Czech Republic}
\maketitle

\section*{Abstract}
We measured the parameter reproducibility and radial electron density profile of capillary discharge waveguides with diameters of \unit[650]{$\mu$m} to \unit[2]{mm} and lengths of \unit[9 to 40]{cm}. To our knowledge, \unit[40]{cm} is the longest discharge capillary plasma waveguide to date. This length is important for \unit[$\geq$10]{GeV} electron energy gain in a single laser driven plasma wakefield acceleration (LPA) stage. Evaluation of waveguide parameter variations showed that their focusing strength was stable and reproducible to \unit[$<0.2$]{\%} and their average on-axis plasma electron density to \unit[$<1$]{\%}. These variations explain only a small fraction of LPA electron bunch variations observed in experiments to date. Measurements of laser pulse centroid oscillations revealed that the radial channel profile rises faster than parabolic and are in excellent agreement with magneto-hydro-dynamic simulation results. We show that the effects of non-parabolic contributions on Gaussian pulse propagation were negligible when the pulse was approximately matched to the channel. However, they affected pulse propagation for a non-matched configuration in which the waveguide was used as a plasma telescope to change the focused laser pulse spot size.


\section{Introduction}
Capillary discharge plasma waveguides~\cite{CDW,CDW3} --also used as active plasma lenses-- allow control over laser pulse diffraction~\cite{LightPipes} and particle bunch divergence~\cite{APL2,APL,APL3,APL4}. They are typically used to guide laser pulses and focus electron bunches, but may also prove useful in a 'plasma telescope' configuration to change laser system focal spot sizes without lengthy transport lines. Optical pulses are focused as a result of the waveguide's radial variation of the plasma electron density~\cite{guidingbasics}; charged particle bunches are focused by the magnetic field associated with the current~\cite{APL}. Capillary plasma waveguides are compact, provide a strong gradient (or focusing)~\cite{comparetoquads} and have a high damage threshold~\cite{highpowerguiding}. They allow for modification of the spot size evolution of intense particle beams and laser pulses and are thus of interest for many applications, including plasma wakefield acceleration~\cite{APLsforPWFA,CDW2,CDLWFA,guidingLWFA}, high harmonic generation \cite{highharmonicgen,HHG1,HHG2,HHG3} and X-ray lasing~\cite{X1,X2}. In this paper we concentrate on waveguide properties that are mainly relevant for laser pulse propagation. Specifically, we measure the transverse electron density profile with sufficient accuracy to discuss the effect of non-parabolic contributions for applications, as well as the stability of the channel shape and position. These parameters are of great importance for the aforementioned applications.

In gas-filled capillary discharge waveguides the plasma is created by a discharge inside a capillary that is initially filled with neutral gas. The current ionizes the gas and heats the plasma. At the same time the plasma is cooled at the capillary walls, leading to a radial variation of the plasma electron density, also called a plasma channel~\cite{plasmachannelsim,plasmachannelsim2}. The channel profile can be expressed as:
\begin{equation}
    \mathrm{n}_{\mathrm{e}}(r) = \mathrm{n}_{\mathrm{e}}(0) + \Delta \mathrm{n}_{\mathrm{e}}(r),
\end{equation} 
where $r$ is the radial coordinate, $\mathrm{n}_{\mathrm{e}}(0)$ is the on-axis plasma electron density and $\Delta \mathrm{n}_{\mathrm{e}}(r)$ represents the shape of the transverse channel profile. The low power guiding properties of the channel are defined by $\Delta \mathrm{n}_{\mathrm{e}}(r)$ and are thus in principal independent of $\mathrm{n}_{\mathrm{e}}(0)$. Like optical fibers for low-power pulses, capillary discharge waveguides have a refractive index profile that is peaked on axis. To match a low-power pulse (laser strength parameter $\mathrm{a}_\mathrm{0}\ll1$) with a transverse Gaussian intensity distribution, a transverse parabolic channel is ideal:
\begin{equation}
    \Delta \mathrm{n}_{\mathrm{e}}(r) = \frac{r^2}{\pi \mathrm{r}_\mathrm{e} w_\mathrm{m}^4},
    \label{eq:parabchannel}
\end{equation}
where $\mathrm{r}_\mathrm{e}$ is the classical electron radius and $w_\mathrm{m}$ is the matched spot size. When the focal laser pulse spot size $w_\mathrm{0}$ equals $w_\mathrm{m}$ and the focus is located at the plasma entrance, the spot size is constant along the channel. If $w_\mathrm{0}$ is different from $w_\mathrm{m}$, the spot size oscillates between $w_\mathrm{0}$ and $w_\mathrm{m}^2/w_\mathrm{0}$ with an oscillation period of $\lambda_{\mathrm{osc}} = \pi^2 w_\mathrm{m}^2/\lambda_\mathrm{l}$, where $\lambda_\mathrm{l}$ is the wavelength of the laser pulse. An offset between the pulse propagation and waveguide axis leads to centroid position oscillations with an oscillation period of $2\lambda_{\mathrm{osc}}$~\cite{reviewpaper}.

There is no current method to measure pulse evolution inside the waveguide; experiments are typically guided by calculations and simulations and require precise knowledge of the channel properties. Radial profiles of capillary discharge waveguides have been measured using longitudinal and transverse diagnostics~\cite{plasmachannelsim,matchedspotsize,previoustransversechannel,previoustransversechannel2}. Although interferometry measurements revealed non-parabolic profiles~\cite{plasmachannelsim,previoustransversechannel}, precision was not sufficient to discuss potential effects on guiding. Higher precision was enabled by centroid oscillation measurements, but did not reveal non-parabolic profiles~\cite{matchedspotsize}. In this paper we reveal for the first time non-parabolic contributions to the channel profile from centroid oscillation measurements. We show that these contributions are negligible when the pulse is approximately matched to the waveguide. However, for non-matched applications (e.g. when waveguides are used in a plasma telescope configuration) these contributions need to be taken into account. 

Applications require reproducible waveguide parameters. For example, the reproducibility of particle bunches accelerated in laser-driven plasma wakefield accelerators (LPAs) is defined by the reproducibility of the wakefield, which is a product of all parts that are involved. Acceleration of high quality bunches requires every component to be well controlled and stable, i.e. the plasma and the drive bunch or pulse. In this paper we characterize the plasma to understand the sources of variations.  We show that the matched spot size $w_\mathrm{m}$ was stable and reproducible to \unit[$0.05$]{\%} for the \unit[9]{cm} long, to \unit[$0.1$]{\%} for the \unit[20]{cm} long and to \unit[$0.2$]{\%} for the \unit[40]{cm} long waveguides. Variations of the longitudinally averaged on-axis plasma electron density were measured to be below \unit[1]{\%}. These results imply that variations caused by a discharge plasma waveguide account for only a small fraction of the parameter variations observed in electron bunches. 

\section{Experimental Setup}
\begin{figure}[htb!]
    \centering
    \includegraphics[width=1\textwidth]{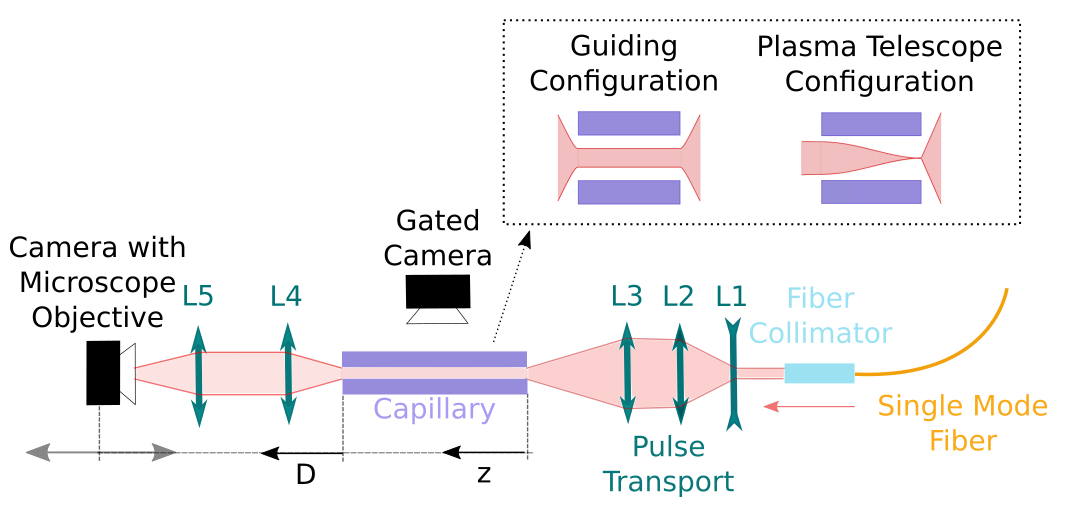}
    \caption{Schematic overview of the experimental setup. The pulse propagates from right to left. The distance along the capillary is $z$ and downstream the capillary is $D$. Optical lenses have the following focal lengths: L$1$: \unit[-10]{cm}, L$2$: \unit[40]{cm}, L$3$: \unit[150]{cm}, L$4$: \unit[100]{cm}, L$5$: \unit[60]{cm}}
    \label{fig:setup}
\end{figure}

Figure~\ref{fig:setup} provides a schematic overview of the experimental setup. We used a laser probe pulse to characterise the plasma channel properties. The pulse was produced by the front end of the BErkeley Lab Laser Accelerator (BELLA) petawatt laser system~\cite{Kei}. Light leaking through a mirror downstream the regenerative amplifier was transported to the experimental area by a single mode optical fiber (see Fig.~\ref{fig:setup}). After transport, the pulse had a central wavelength of \unit[$\lambda_\mathrm{l} = $780]{nm}, a pulse length of \unit[$\tau$ $\simeq$600]{ps} and a focal point spot size adjustable from \unit[$\sim$70 to 105]{$\mu$m}. The pulse energy was \unit[$<$4]{nJ} which is low enough to not modify the channel profile. The capillary was discharged and measurements were obtained at a repetition rate of \unit[1]{Hz}. The pulse was set to arrive \unit[$\sim$0 to 400]{ns} after the peak of the discharge current pulse.  

The transverse, time-integrated laser pulse profile was imaged with a charged-coupled-device (CCD) camera, as shown in Fig.~\ref{fig:setup}. The camera had 1626 x 1236 pixels and a resolution of \unit[0.935]{$\mu$m/pixel}. The camera was mounted onto a translation stage; the optical setup allowed changing the imaging location from \unit[$\sim$30]{cm} upstream the capillary entrance to \unit[$\sim$40]{cm} downstream the capillary exit. Figure~\ref{fig:beamsizeinout} (left) shows an example of a measured (single-shot, background subtracted) transverse laser pulse intensity distribution at the focal point location (which was also the location of the capillary entrance aperture).

The plasma was created by a discharge current in a gas-filled capillary. The capillary bulk material was either Sapphire (Al$_2$O$_3$) or a machinable glass ceramic (MGC). Sapphire is transparent and allows for transverse plasma light diagnostics~\cite{uniformitychannel}; the MGC is opaque but is cheaper, can be made longer in a single piece and is easily mechanically machinable. Two gas inlets (located \unit[6]{mm} from the upstream and downstream capillary ends) allowed filling the capillaries with Hydrogen gas. Both capillary ends were open such that gas escapes to the vacuum chamber. Table~\ref{tab:capparameters} lists the capillaries that were used for this paper. We chose parameters that are either relevant for laser driven plasma wakefield acceleration (\#1, \#2, \#4) or plasma telescopes (\#3).

\begin{table}[htb!]
\begin{center}
\begin{tabular}{|c|l|c|c|l|l|l}
\cline{1-6}
Cap. & Material & Diameter {[}$\mu$m{]} & Length {[}cm{]} &  $n_{\mathrm{0i}}$ [atoms/cm$^3$] & $w_\mathrm{m}$ [$\mu$m] \\ \cline{1-6}
\#1   &    Sapphire     &       650 $\pm$ 30           &        9         & (1.6-13)$\times$10$^{17}$& 105-75& \\ 
\#2   &    Sapphire     &       830  $\pm$ 50          &        20        &  (0.7-6.6)$\times$10$^{17}$& 180-115& \\
\#3   &      MGC      &       2000  $\pm$ 20         &        20        &   (0.65-2.4)$\times$10$^{17}$& $\sim$650-235& \\
\#4   &      MGC      &       830  $\pm$ 20          &        40        &   (1.5-6.6)$\times$10$^{17}$& 180-115& \\ \cline{1-6}
\end{tabular}
\caption{Overview of capillary parameters and operation range.}
\label{tab:capparameters}
\end{center}
\end{table}

Electrodes were located at each end of the capillary. Switching capacitors charged to \unit[$\sim$20]{kV} induced a gas-breakdown. The current pulse profile was close to sinusoidal with a rise time between 340 and \unit[420]{ns} and a peak amplitude between 520 and \unit[620]{A} (depending on the capillary length and radius, and initial neutral gas pressure). We note that plasma channel properties are only weakly dependent on discharge current pulse length and peak current~\cite{plasmachannelsim2}. The values for the on-axis plasma electron densities $\mathrm{n}_{\mathrm{e}}(0)$ are known from measurements of the initial neutral gas density that were compared to previous plasma electron densities~\cite{GVD} and range from \unit[(0.1 to 50)$\times10^{17}$]{electrons/cm$^{3}$}. The maximum available pressure was limited by the speed of the vacuum pumping system and the lowest available pressure was limited by the timing reproducibility of the discharge. For the sapphire capillaries, spectrally resolved plasma light images suggest that ablation from the capillary walls is negligible for the parameters in Tab.~\ref{tab:capparameters}. Aluminum and oxygen spectral lines were observed for lower initial neutral gas densities than Tab.~\ref{tab:capparameters} where the plasma temperature is higher~\cite{khzTony}.

As summarized in Tab.~\ref{tab:capparameters}, plasma waveguides with 9 to \unit[40]{cm} length were analyzed. Producing shorter channels is typically straightforward. To our knowledge, \unit[40]{cm} is the longest discharge capillary waveguide that has been demonstrated. In the context of laser driven plasma wakefield acceleration, reaching \unit[10]{GeV} electron energy gain in a single stage requires an on-axis plasma electron density lower than \unit[$\mathrm{n}_{\mathrm{e}}(0)=\,$2.8$\times10^{17}$]{electrons/cm$^3$}, based on 1-D estimates~\cite{reviewpaper}. At \unit[$\mathrm{n}_{\mathrm{e}}(0)=\,$2.8$\times10^{17}$]{electrons/cm$^3$} the dephasing length is \unit[$L_d\sim$40]{cm} for a \unit[800]{nm} driver. Even longer channels are possible, but may require timing jitter mitigation techniques~\cite{trig1,doublecurrentpulse,lasertrigger,lasertrigger2}.

Capillary~\#3 has a diameter of \unit[2000]{um}, which is more than double of what is typically used for guiding. We tested this capillary to demonstrate the 'plasma telescope' concept (see Fig.~\ref{fig:setup}), a compact way to modify laser pulse spot size. Large diameter capillaries are also interesting for particle bunch transport, where the capillary radius often limits particle acceptance or the maximum bunch size. Larger bunch sizes (and therefore lower bunch intensities) are desired to minimize wakefield effects in active plasma lenses. Even larger diameters are possible, but require an increase in the speed of the vacuum pumping system.

\section{Experimental Results}
\subsection{Measurement of the Matched Spot Size}
\label{sec:wm}
\begin{figure}[htb!]
    \centering
    \includegraphics[width=0.8\textwidth]{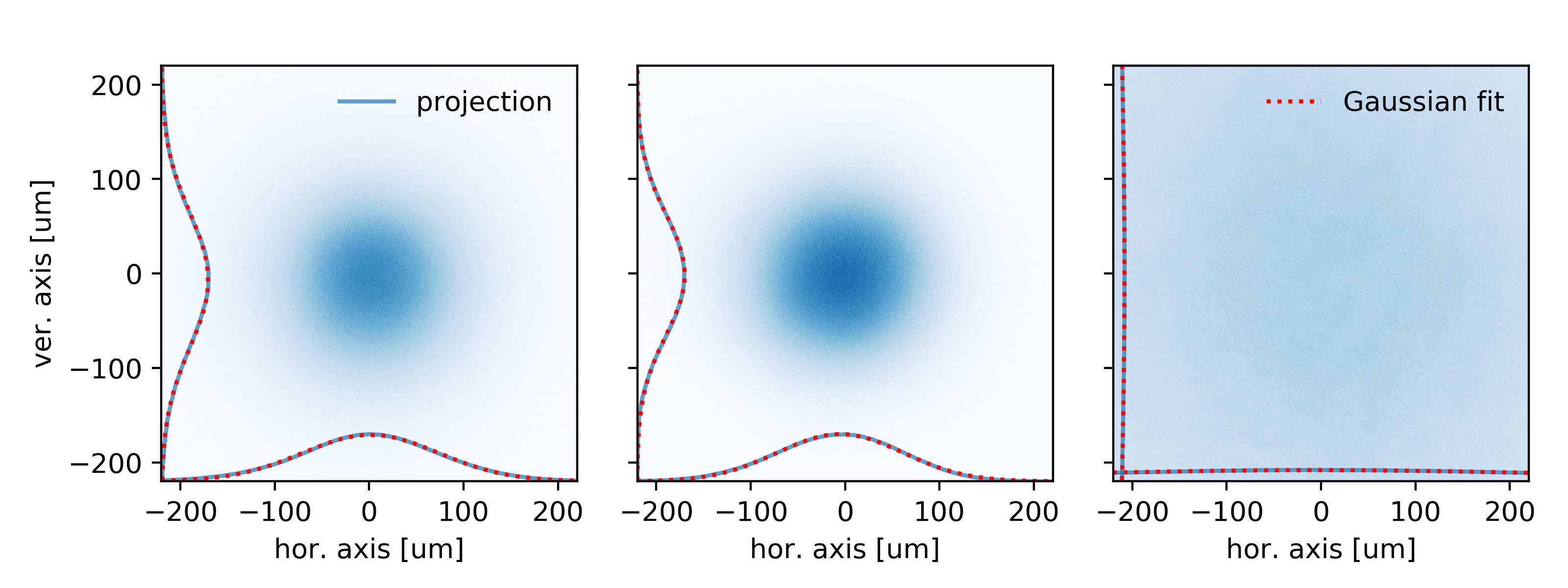}
    \caption{Examples of experimentally measured transverse pulse intensity distributions. Left: at vacuum focus; Center: at the the capillary exit plane for when the pulse propagated in the \unit[40]{cm} long capillary~\#4; Right: at the capillary exit plane when the pulse propagated in vacuum. Blue lines show the horizontal and vertical projections of the camera images. Red dotted lines show the results of Gaussian fits of the projections. The left and middle plots are on the same linear color-scale; the color-scale of the right plot is enhanced by a factor of ten.}
    \label{fig:beamsizeinout}
\end{figure}

Figure~\ref{fig:beamsizeinout} shows examples of transverse pulse intensity distributions measured when using the \unit[40]{cm} long capillary~\#4: the left image shows the pulse at the capillary entrance (which is also the location of the focal point); the right image shows the central section of the pulse after \unit[40]{cm} of vacuum propagation (root-mean-square [rms] spot size \unit[$\sim300$]{$\mu$m}); the center image shows an example of the pulse in the capillary exit plane after transport in the plasma waveguide. The transverse intensity distribution of the guided pulse was similar to the input pulse.

We estimated the waveguide energy transmission $E_\mathrm{t}$ by comparing the total counts of the background subtracted pulse input and output camera images. Within the error of this measurement (\unit[$\Delta E_\mathrm{t}\sim$2]{\%}) and for reasonable experimental settings, the energy transmission was approximately \unit[100]{\%}. All energy that enters the waveguide, exits. Thus, pulse scattering off the plasma is negligible for waveguides with a density-length product of up to at least \unit[6$\times10^{18}$]{atoms/cm$^{3}\times$cm}.

Plasma channels focus laser pulses due to a radially varying refractive index. The profile is defined by the radial plasma electron density distribution $\Delta \mathrm{n}_{\mathrm{e}} (r)$, which is defined by thermal effects and is closely tied to the capillary radius. The smaller the radius, the steeper the temperature gradient, the stronger the focusing strength of the channel. The matched spot size $w_\mathrm{m}$ can be obtained from the radial electron density distribution according to Eq.~\ref{eq:defrm}~\cite{rmfromnpe}:
\begin{equation}
    8 \pi \mathrm{r}_\mathrm{e} \int_0^{\infty} \left[\mathrm{n}_{\mathrm{e}}(r) \left(\frac{2r^2}{\mathrm{w}_\mathrm{m}^2}-1\right) e^{-\frac{2r^2}{\mathrm{w}_\mathrm{m}^2}} \right] r\,dr -1=0,
    \label{eq:defrm}
\end{equation}
where $r$ is the radial coordinate. The spot size evolution of a transversely Gaussian, low-intensity pulse that is focused at the entrance of a parabolic channel is described by Eq.~\ref{eq:spotsizeevolution}~\cite{matchedspotsize}:

\begin{equation}
    w^2 (z) = \frac{w_\mathrm{i}^2}{2} \Big[1+\frac{w_\mathrm{m}^4}{w_\mathrm{i}^4}+ \left( 1- \frac{w_\mathrm{m}^4}{w_\mathrm{i}^4} \right) \mathrm{cos}\left(\frac{2\pi z}{\lambda_{\mathrm{osc}}} \right) \Big],
    \label{eq:spotsizeevolution}
\end{equation}
where $w$ is the transverse spot size at the location $z$ along the waveguide, $w_\mathrm{i}$ is the input spot size and $\lambda_{\mathrm{osc}}$ is the spot size oscillation length. Unless the pulse propagation axis overlaps with the channel axis, the pulse centroid position oscillates along the channel. The pulse centroid position at the capillary exit $O_{\mathrm{pulse}}$ (with respect to the incoming laser axis) is defined by the trajectory of the pulse centroid along the channel. When the transverse channel profile is parabolic and the incident low-power pulse parallel to the waveguide axis, the centroid oscillations can be described by~\cite{matchedspotsize}:

\begin{equation}
    x_c = x_i\,\mathrm{cos}\left(\frac{4\pi z}{\lambda_{\mathrm{osc}}}\right)
    \label{eq:centroid}
\end{equation}
where $x_i$ is the offset between the pulse propagation and waveguide axis. 

Measuring the pulse centroid position at the capillary exit $O_{\mathrm{pulse}}$ as a function of the offset between the pulse and capillary axis  $O_{\mathrm{cap}}$ (defined by the translation of the capillary axis with respect to the incoming laser axis) allows for evaluation of the matched spot size of a parabolic channel. The technique is described in~\cite{matchedspotsize}. If the channel is parabolic, the measured relation is linear and the matched spot size can be calculated from Eq.~\ref{eq:rmreconst}:

\begin{equation}
w_\mathrm{m} = \sqrt{\frac{2z}{k[\mathrm{cos}^{-1} (d O_{\mathrm{pulse}}/d O_{\mathrm{cap}})+2\pi j]}}, 
\label{eq:rmreconst}
\end{equation}
where $dO_{\mathrm{pulse}}/dO_{\mathrm{cap}}$ is the near-axis derivative of $O_{\mathrm{pulse}}$ versus $O_{\mathrm{cap}}$ and $k$ is the laser wavenumber. In this work the integer $j$ was determined by measuring the response of the laser pulse to the initial neutral gas density.

\begin{figure}[htb!]
    \centering
    \includegraphics[width=0.80\textwidth]{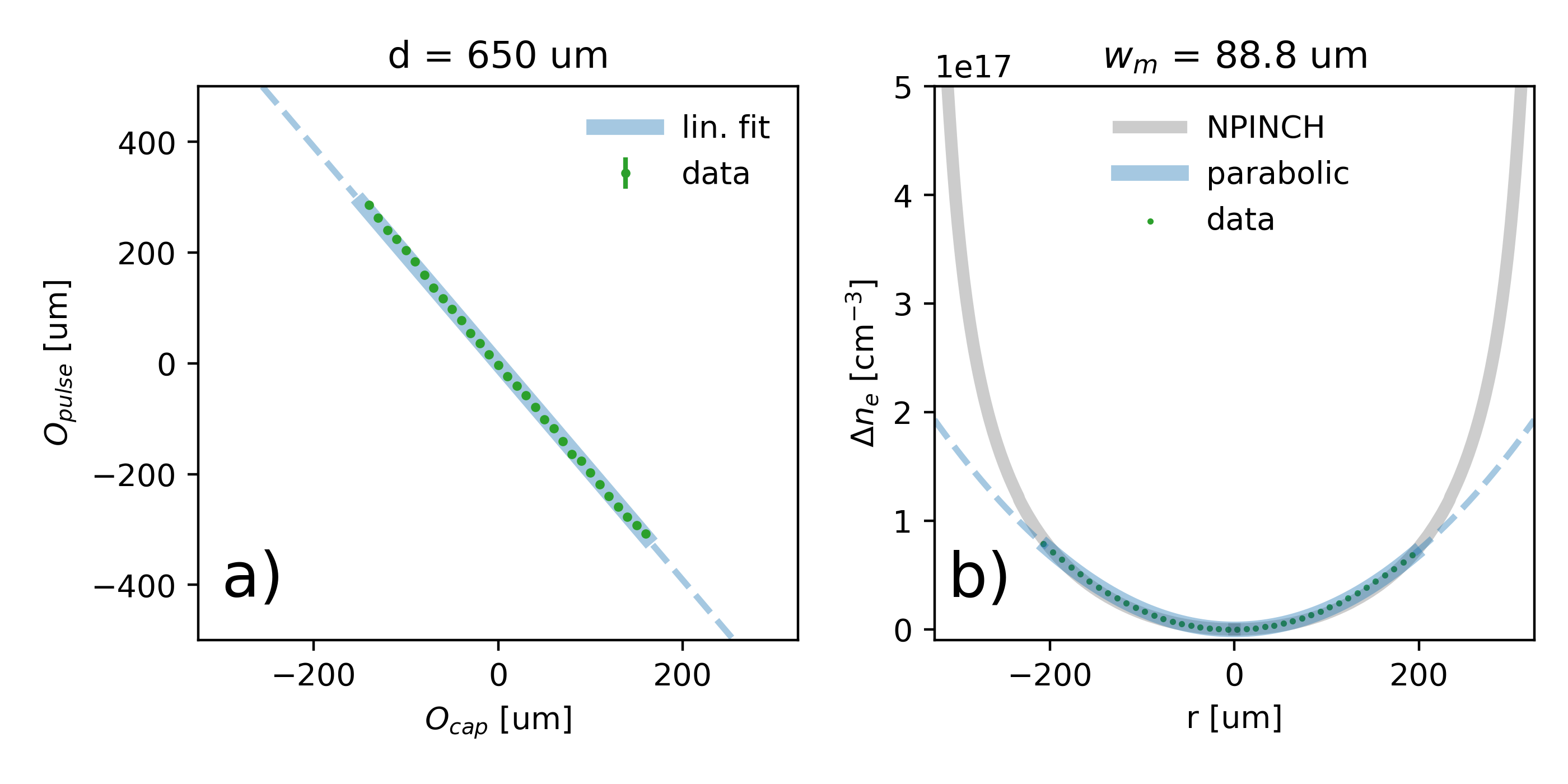}
    \caption{Reconstructed radial plasma electron density profile. a) Measurements of the centroid position of the guided pulse at the capillary exit $O_{\mathrm{pulse}}$ as a function of the parallel capillary offset with respect to the laser propagation axis $O_{\mathrm{cap}}$ (green markers). Error-bars (standard deviation of the individual measurements) are not visible since they are smaller than the marker size. The blue line shows a linear fit to the data; it is solid over the measurement range and dashed outside (assuming the continuation of a parabolic channel outside the measurement range). b) calculated relative change of the radial plasma electron density $\Delta \mathrm{n}_{\mathrm{e}}(r)$ (green markers) compared to 1D NPINCH simulation results (gray line). The blue lines corresponds to the result of the fit shown in panel a).}
    \label{fig:parabchannel}
\end{figure}

Figure~\ref{fig:parabchannel}a shows the measured relation of $O_{\mathrm{pulse}}$ as a function of $O_{\mathrm{cap}}$ for capillary~\#1 and an initial neutral gas density number of \unit[$n_{i0} = 6.5\times10^{17}$]{atoms/cm$^3$}. The pulse centroid position was defined from Gaussian fits to the projections of the measured pulse intensity distributions. Measurement ranges were limited to \unit[$|O_{\mathrm{cap}}| \lesssim 200$]{$\mu$m} as for larger values of $O_{\mathrm{cap}}$ the pulse interacted with the capillary walls.

In Fig.~\ref{fig:parabchannel}a, $O_{\mathrm{pulse}}$ was a linear function of $O_{\mathrm{cap}}$ over the measurement range. The result is consistent with the pulse experiencing a parabolic channel profile ($\Delta \mathrm{n}_{\mathrm{e}}(r)$ according to Eq.~\ref{eq:parabchannel}). We then used Eq.~\ref{eq:rmreconst} to calculate $w_\mathrm{m}$ and show the resulting radial distribution of the plasma electron density in Fig.~\ref{fig:parabchannel}b (blue line). Measurements yielded excellent agreement with 1D Magneto-Hydro-Dynamic (MHD) simulations results (gray line). The simulations were performed with the 1D code NPINCH~\cite{NPINCH}, using the experimental capillary and atomic density parameters as input. These simulations are consistent with previous theoretical and simulation work~\cite{plasmachannelsim} showing that capillary discharge waveguides do not produce parabolic density profiles over the entire capillary transverse extent. However, over a sufficiently small radius (\unit[$|O_{\mathrm{cap}}| \gtrsim 200$]{$\mu$m}), a parabolic assumption is reasonable (see Fig.~\ref{fig:parabchannel}b)).

We performed horizontal and vertical centroid oscillation measurements for all capillaries over the accessible density range. Values for $w_\mathrm{m}$ are summarized in Tab.~\ref{tab:capparameters}~\cite{exptimint}. The matched spot size obtained from horizontal and vertical measurements agreed within \unit[2]{\%}. Additionally, when the pulse propagation axis was approximately aligned to the waveguide axis, the guided pulse intensity distributions were radially symmetric (for all tested capillaries, see e.g. Figs.~\ref{fig:beamsizeinout} and \ref{fig:resize}c) and the horizontal and vertical pulse sizes agreed to better than \unit[$<1$]{$\mu$m} (or \unit[1.4]{\%}) for the \unit[9]{cm} long, \unit[$<3$]{$\mu$m} (or \unit[2.7]{\%}) for the \unit[20]{cm} long,  \unit[$<3$]{$\mu$m} (or \unit[2.7]{\%}) for the \unit[40]{cm} long waveguides. Therefore, the matched spot size was symmetric to better than \unit[3]{\%}.

\subsection{Waveguide Parameter Reproducibility}
\label{sec:length}

Applications and multi-shot measurements require channel properties to be stable and reproducible from discharge to discharge. Parameters such as the channel central-axis, the focusing strength, or the on-axis plasma electron density are demanded to be repeatable from event to event and stable over long timescales. For example, variations in the matched spot size change the pulse evolution inside the waveguide as well as the pulse divergence downstream. To evaluate waveguide parameter variations, we observed their effects on the probe pulse. We also decided to use a combination of capillary length $L$ and matched spot size $w_\mathrm{m}$ for which: 

\begin{enumerate}
    \item the transverse spot size at the capillary exit was equal to the pulse vacuum focal spot size; the spot underwent an integer number $n$ of oscillations,  $L = n \lambda_{\mathrm{osc}}$. Analysis of a similar sized input and guided pulse facilitated the interpretation of the results; 
    \item the pulse centroid position underwent approximately half an integer number of oscillations, $L = (n/2) \lambda_{\mathrm{osc}}$, such that the pulse centroid offset at the capillary exit plane was maximised and measurement sensitivity optimized.
\end{enumerate} 

We chose capillary~\#1 (\unit[9]{cm} length~\cite{beamsfromdischargecap2}) together with an initial neutral gas density of \unit[$n_{i0} = 6.5\times10^{17}$]{atoms/cm$^3$} that resulted in a matched spot size of \unit[$\mathrm{w}_\mathrm{m}=$88.8]{$\mu$m} (see Fig.~\ref{fig:parabchannel}). For the \unit[780]{nm} probe pulse, the theoretical spot size oscillation length was \unit[$\lambda_{\mathrm{osc}}$=9.9]{cm} and the centroid oscillation length was \unit[$2\lambda_{\mathrm{osc}}$=19.8]{cm}. The pulse therefore underwent approximately half a centroid position oscillation and a full spot size oscillation along the capillary.

\begin{figure}[htb!]
    \centering
    \includegraphics[width=1\textwidth]{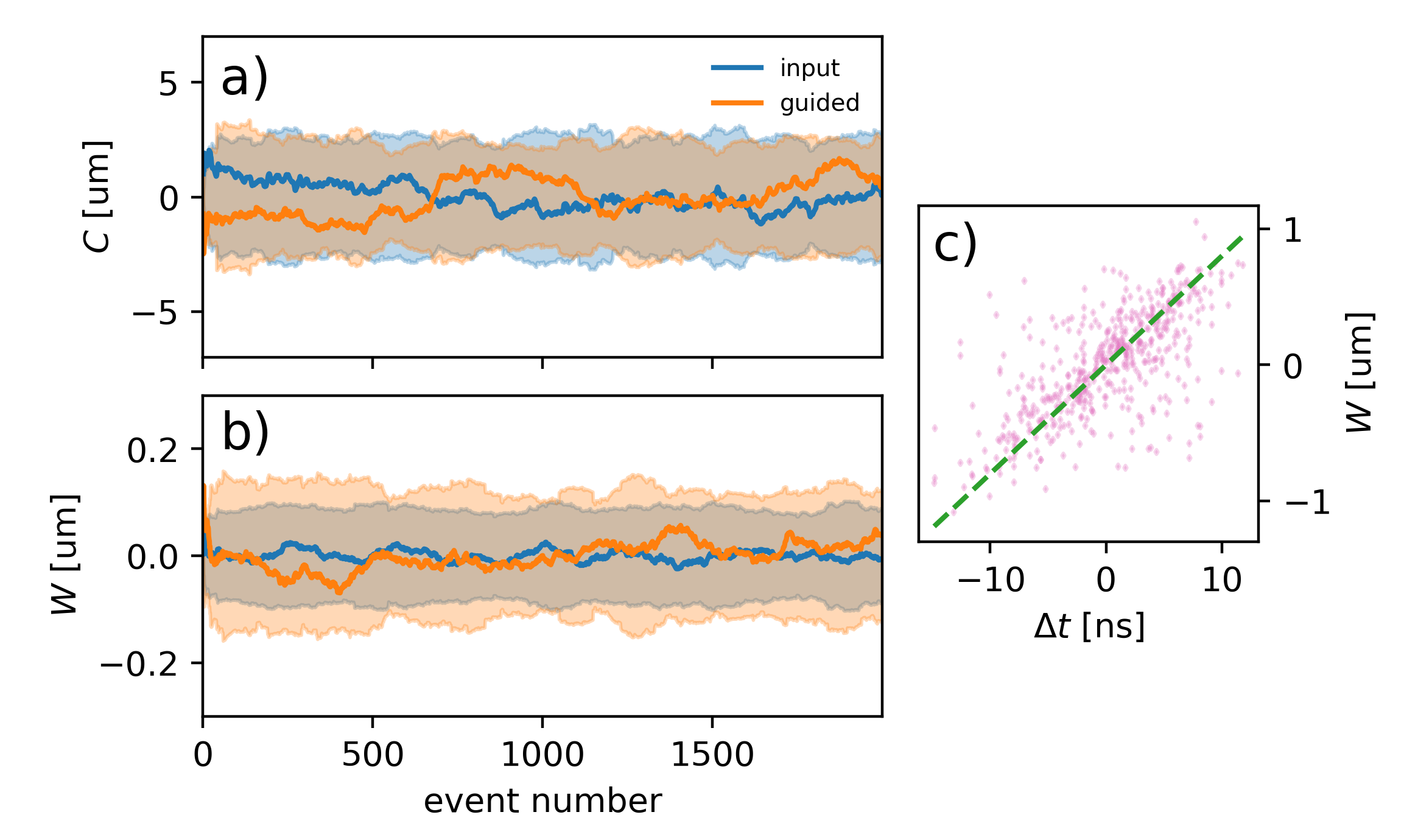}
    \caption{a) Pulse centroid position variations $C = c-\bar{c}$, where $c$ is the measured pulse center and $\bar{c}$ the average value of all measurements; b) spot size deviations $W = w-\bar{w}$ , where $w$ is the rms pulse size and $\bar{w}$ the average value of all measurements. The Figure shows the running average of 100 measurements (solid line) as well as their standard deviation (error-band) for capillary~\#1. Measurements for the incoming pulse at the capillary entrance are shown in blue and the ones for the guided pulse at the capillary exit in orange. c) Correlation between $\Delta w$ and the timing jitter between the discharge and the probe pulse ($\Delta t$) for an initial neutral gas density of  \unit[$n_{i0} = 6.5\times10^{17}$]{atoms/cm$^3$} for capillary~\#2. The green dashed line shows the result of a linear fit.}
    \label{fig:stability}
\end{figure}

Fig.~\ref{fig:stability}a compares variations of the guided pulse centroid position at the capillary exit (orange) to the ones of the incoming pulse at the capillary entrance (blue). The pulse centroid was defined as the location of the peak of the Gaussian fit to the projections. Deviations from the centroid reference $c$ are the quadratic sums of the horizontal $c_\mathrm{x}$ and vertical $c_\mathrm{y}$ contributions ($c=\sqrt{c_\mathrm{x}^2+ c_\mathrm{y}^2}$). The top plot of Fig.~\ref{fig:stability} compares the running average of 100 measured pulse centroid deviations $C = c-\bar{c}$, where the mean centroid position $\bar{c}$ was subtracted from the individual measurement $c$; the bottom plot of Fig.~\ref{fig:stability} compares running average of 100 pulse spot size deviations $W = w - \bar{w}$, where the mean spot size \unit[$\bar{w}=(74\pm1$)]{$\mu$m} was subtracted from the individual measurement $w$. Respectively colored error-bands show the running average of the standard deviations of 100 measurements.

From the data presented in Fig.~\ref{fig:stability}a we calculated the standard deviation of all measurements. Variations of the incoming pulse \unit[$\Delta C$ = 2.7]{$\mu$m} are rms and were equal to the ones of the guided pulse. This means that within the resolution of the measurement (\unit[$1$]{$\mu$m}), the central position of the created plasma channel was repeatable and stable over the measured timescale. The overall length of the measurements was limited to \unit[$\sim$30]{minutes} by a thermal drift of the incoming probe pulse, which could be mitigated in the future with active pointing feedback.

The rms spot size variation $\Delta W$ (Fig.~\ref{fig:stability}b) increased after guiding by \unit[$\Delta W$ = 0.04]{$\mu$m} (see bottom plot of Fig.~\ref{fig:stability}).  Spot-size variations are a result of the variations of the matched spot size. The measured values translate to variations of $w_{\mathrm{m}}$ around \unit[0.05]{\%}. This increase in $\Delta W$ was measurable, but negligible with respect to the absolute spot size of \unit[$\bar{w}$=74]{$\mu$m} and translates to pulse peak intensity variations $\Delta I$ of \unit[$\sim$0.1]{\%} for $w_\mathrm{0}=w_\mathrm{m}$.

%


We performed the same set of measurements for capillaries~\#2 and \#4. There was no measurable increase in $\Delta C$. Spot size variations increased by \unit[$\Delta W$ = 0.26]{$\mu$m} for capillary~\#2 and \unit[$\Delta W$ = 0.52]{$\mu$m} for capillary~\#4, corresponding to \unit[0.5]{\%} and \unit[1]{\%} in pulse peak intensity variations (when $w_\mathrm{0}=w_\mathrm{m}$). They translate to variations in matched spot size that are around \unit[0.1]{\%} for capillary~\#2 and \unit[0.2]{\%} for capillary~\#4. 

We found that for capillaries longer than \unit[20]{cm}, part of the increase in spot size variations could be explained by timing jitter between the discharge and the laser pulse arrival time ($\Delta t$). Experiments showed that the amplitude of the timing jitter depends on the initial neutral gas density ($n_{i0}$), the capillary length and radius as well as the discharge voltage. For example, for \unit[$n_{i0} = 2\times10^{17}$]{atoms/cm$^3$} the rms value of the timing jitter was \unit[$\Delta t \sim $5]{ns} for the \unit[20]{cm} long capillary~\#2, \unit[$\Delta t = $12]{ns} for the \unit[2]{mm} diameter capillary~\#3  and \unit[$\Delta t = $9]{ns} for the \unit[40]{cm} long capillary~\#4. Figure~\ref{fig:stability}c shows the correlation measurement for capillary~\#2 for \unit[$n_{i0} = 2\times10^{17}$]{atoms/cm$^3$} as well as the result of a linear fit to the measurements (green dashed line). Improving the discharge timing jitter~\cite{trig1} by e.g. using an additional current pulse~\cite{doublecurrentpulse} to create the plasma, operating at a higher discharge voltage or using a laser trigger~\cite{lasertrigger,lasertrigger2}) could decrease variations by e.g. \unit[$\Delta W$ = 0.11]{$\mu$m} (or \unit[-20]{\%}) in the case of capillary~\#2. The correlation was not measurable with capillary~\#1 as jitter values were below \unit[$\Delta t < $1]{ns}. 

For our laser pulse input parameters and the minimum matched spot size of capillary~\#3 (\unit[$\mathrm{w}_\mathrm{m}>230$]{$\mu$m}) the pulse at the capillary exit was always much larger than the input and showed significant higher order mode contributions (see Sec.~\ref{sec:plasmatelescope}). This resulted in limited accuracy in the measurement of the centroid (see condition 2. above) and spot size variations and results are therefore not presented. In the future the laser angle exiting the waveguide could be measured to increase accuracy.

\subsection{Observation of Higher Order Contributions to the Transverse Channel Profile}
\label{sec:nonlinearityinr}

\begin{figure}[htb!]
    \centering
    \includegraphics[width=0.8\textwidth]{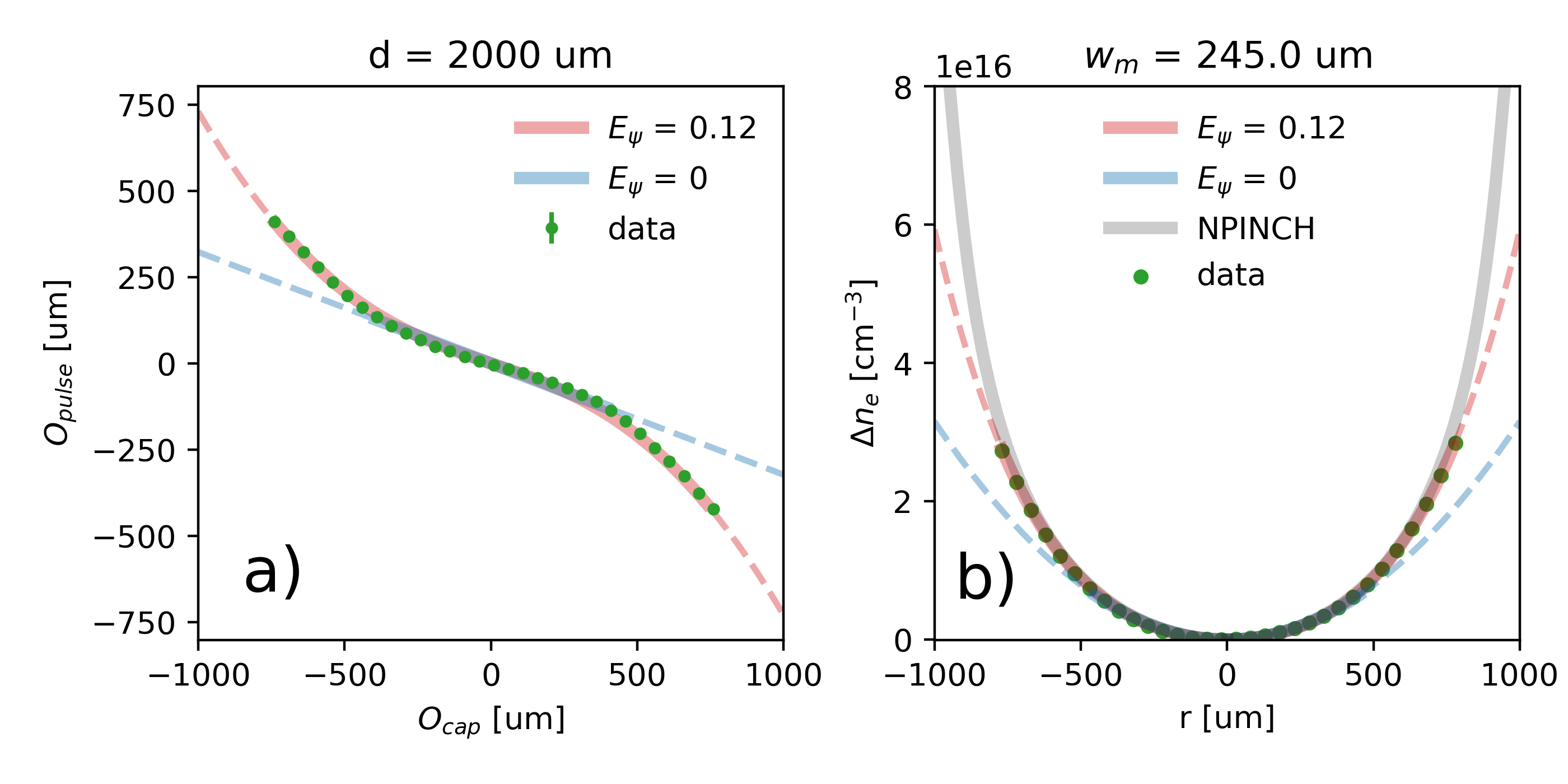}
    \caption{Reconstructed radial plasma electron density profile: a) measurements of the centroid position of the guided pulse at the capillary exit $O_{\mathrm{pulse}}$ (green markers) as a function of the parallel capillary offset with respect to the laser propagation axis $O_{\mathrm{cap}}$. Error-bars (standard deviation of the individual measurements) are not visible as they are smaller than the marker size. The blue and red solid (within the measurement range) and dashed (outside the measurement range) lines show the calculated centroid oscillation scan result corresponding to the same colored density profiles on the right.  b) calculated relative change of the plasma electron density $\Delta \mathrm{n}_{\mathrm{e}}$ as a function of radial position $r$ for a parabolic channel (blue line), a channel with a $r^4$ component (red line) compared to the result from to NPINCH simulations (gray line).}
    \label{fig:hextranschannel}
\end{figure}

The excellent reproducibility of the waveguide is a prerequisite for applications, e.g. to produce stable electron bunches from a LPA. Together with a reproducible probe pulse it also enabled measurements of the transverse channel profile with unprecedented precision. Figure~\ref{fig:hextranschannel}b shows a centroid oscillation measurement for the \unit[2000]{$\mu$m} diameter capillary~\#3 (green dots). The alignment offset was limited to \unit[$|O_{\mathrm{cap}}| \leq\sim700$]{$\mu$m}; outside this range the laser pulse interacted with the capillary walls and accurate identification of the pulse centroid position was no longer possible. For \unit[$|O_{\mathrm{cap}}|\gtrsim500$]{$\mu$m}, the guided pulse intensity distribution deviated from Gaussian; the pulse centroid position was therefore calculated from the center-of-mass.

The blue line in Fig.~\ref{fig:hextranschannel}a shows a linear fit (according to Eq.~\ref{eq:rmreconst}) to the centroid oscillation result. It does not provide a good description of the measurement for \unit[$|O_\mathrm{cap}|>400$]{um}. To find a model that can describe the results better, we calculated the centroid position response to different radial electron density distributions $\mathrm{n}_{\mathrm{e}}(r)$. We found that Eq.~\ref{eq:channel} provides an excellent description of the measurements in general, as was predicted in~\cite{rmfromnpe}: 

\begin{equation}
\mathrm{n}_{\mathrm{e}}(r) = n_\mathrm{e}(0) + \frac{(1-E_{\psi})}{\pi \mathrm{r}_\mathrm{e} w_\mathrm{m}^4} r^2 + \frac{ E_{\psi}}{2 \pi \mathrm{r}_\mathrm{e} w_\mathrm{m}^6} r^4,
\label{eq:channel}
\end{equation}
where $E_{\psi}$ is a scaling factor for the relative contributions of the $r^2$ and $r^4$ terms. When $E_{\psi}=0$, the radial channel profile is parabolic, equal to Eq.~\ref{eq:parabchannel}. 

Figure~\ref{fig:hextranschannel}a shows that adding an $r^4$ ($E_{\psi}>0$) component to the radial plasma electron density profile (red line) according to Eq.~\ref{eq:channel} allowed to greatly improve the agreement between the model ($E_{\psi}$ and $w_\mathrm{m}$ are free parameters) and the measurement. We calculated the result of centroid oscillation scans for different parameters of $w_\mathrm{m}$ and E$_\psi$ until the average deviation between measurements and the model was minimized. Figure~\ref{fig:hextranschannel}a shows the relative radial plasma electron densities that correspond to the models shown in Fig.~\ref{fig:hextranschannel}b. A comparison with NPINCH simulation results (gray line) confirmed that the channel profile rises stronger than parabolic (blue line, $E_{\psi}=0$). Simulation results also indicate that a description of the channel profile close to the capillary walls (\unit[$|r|\gtrsim 700$]{$\mu$m}, outside the measured range) may require adding additional higher order terms to Eq.~\ref{eq:channel}.

Figures~\ref{fig:parabchannel}b and \ref{fig:hextranschannel}b show that the central regions ($|r|\lesssim 2 w_\mathrm{m}$) of the measured channel profiles were well described by Eq.~\ref{eq:parabchannel}. For $|r| > 2 w_\mathrm{m}$, the $r^4$ contribution became significant, with increasing values of $E_{\psi}$ for larger capillary diameters and higher initial neutral gas pressures. Values of $E_{\psi}$ varied only slightly with discharge timing and initial neutral gas density. Over our measurement range (see Tab.~\ref{tab:capparameters}) they range from $0.05<E_{\psi}<0.15$ for capillaries~\#2 and \#4, from $0.1<E_{\psi}<0.2$ for capillary~\#3. For capillary~\#1 they were generally small (below $E_{\psi}<0.05$). Outside the measurement ranges (see red and blue dashed lines in Figs.~\ref{fig:parabchannel}b and \ref{fig:hextranschannel}b), there is a discrepancy between the models and the NPINCH simulation result. Polynomial fits to NPINCH simulation results revealed that all capillary diameters require an additional $r^6$ term to obtain a good description of the density profile close to the capillary walls. 

\section{Implications of Parameter Reproducibility \& Non-Parabolic Channel Contributions for Applications}
\subsection{Laser Driven Plasma Wakefield Accelerators}

\begin{figure}[htb!]
    \centering
    \includegraphics[width=0.5\textwidth]{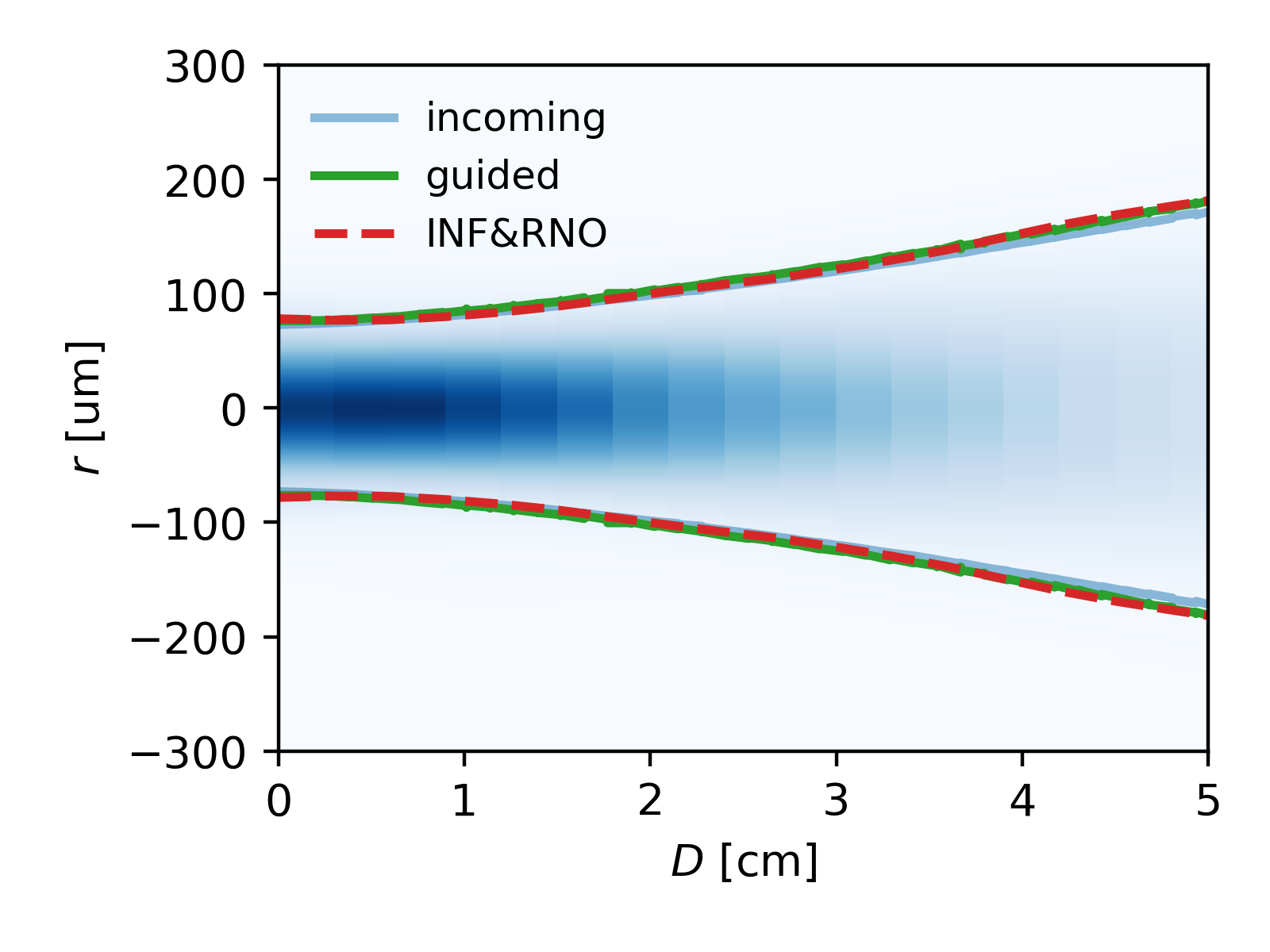}
    \caption{Waterfall plot of the simulated pulse intensity evolution downstream the capillary exit $D$. The red dashed line shows the rms spot size from Gaussian fits to the intensity distribution from the simulations for a parabolic channel using \unit[$w_\mathrm{m}=88.8$]{$\mu$m} (red). The green line shows the measured evolution of the rms size of the pulse after guiding. The blue line shows the pulse evolution downstream the vacuum focus.}
    \label{fig:divergence}
\end{figure}

When using waveguides in the context of laser-driven plasma wakefield acceleration the matched spot size $w_\mathrm{m}$ is typically chosen be close to the pulse focal spot size $w_\mathrm{0}$. The pulse is contained to the parabolic region ($w\sim w_\mathrm{m}$) of the channel. For these circumstances, as shown in  Fig.~\ref{fig:divergence}, we experimentally observed that the pulse evolution downstream the capillary exit (green solid line) was approximately equal to the pulse evolution downstream vacuum focus (blue solid line). There is good agreement between the two measurements, over approximately two Rayleigh ranges (\unit[$Z_\mathrm{r}\sim$2.2]{cm}), which shows that pulse quality was not degraded.

To confirm that the radial profile measured in Sec.~\ref{sec:wm} is consistent with the evolution of the pulse exiting the waveguide, propagation simulations were performed. The pulse mode was retrieved from a Gerchberg-Sexton algorithm on the measured fluence evolution along the propagation axis $z$. We performed INF\&RNO~\cite{inferno1,inferno2} calculations using this mode as input. The plasma channel was modelled according to the results of Fig.~\ref{fig:parabchannel}, parabolic with  \unit[$w_\mathrm{m}$=88.8]{$\mu$m}. Along the plasma, the density profile was uniform with linear density ramps to vacuum at each end. Figure~\ref{fig:divergence} shows the simulated pulse intensity evolution downstream the capillary exit ($D=0$). On top of the waterfall plot of the pulse intensity, we compare the results of Gaussian fits on the simulated (red dashed line) and measured (green solid line) data. The good agreement between the two curves confirmed the experimentally measured channel profile from Fig.~\ref{fig:parabchannel}.


In Sec.~\ref{sec:length} we demonstrated the excellent reproducibility of the waveguide parameters that control pulse guiding (second term on the right hand side of Eq.~\ref{eq:parabchannel}). For capillary~\#2, the matched spot size was repeatable to \unit[0.1]{\%}. Changing the matched spot size by \unit[$\sim$0.1]{\%}, for example from \unit[88.8]{$\mu$m} to \unit[88.9]{$\mu$m} changes the pulse peak intensity of a Gaussian pulse with a spot size of \unit[$w_0=$88.8]{$\mu$m} by \unit[0.25]{\%}. These variations translate to a change in the laser strength parameter $a_\mathrm{0}$ of \unit[0.12]{\%}. The pulse peak intensity variations caused by the waveguide are therefore much smaller than typical intensity variations (\unit[$\sim10$]{\%}) at the focal plane of petawatt class laser systems~\cite{Kei}.

LPAs additionally require reproducibility of the on-axis plasma electron density $\mathrm{n}_{\mathrm{e}}(0)$ (first term on the right hand side of Eq.~\ref{eq:parabchannel}): plasma electron density variations result e.g. in variations of the accelerating gradient as the wakefield phase and amplitude changes. We evaluated variations of $\mathrm{n}_{\mathrm{e}}(0)$ by using a Group Velocity Dispersion (GVD) technique described in~\cite{GVD}. Longitudinal average plasma density values are obtained by tracking the phase of interference fringes. Measurements were performed on capillary~\#2 on the BELLA petawatt experimental setup~\cite{Kei,BELLA} for an on-axis density of \unit[$\mathrm{n}_{\mathrm{e}}=(2\times10^{17})$]{electrons/cm$^3$} and were \unit[$<$1]{\%} (with accuracy limited by probe pulse fluctuations). At \unit[$\mathrm{n}_{\mathrm{e}}=(2\times10^{17})$]{electrons/cm$^3$} the peak accelerating field (estimated based on the cold plasma wave-breaking limit) is \unit[43]{GV/m}. A \unit[1]{\%} variation in $\mathrm{n}_{\mathrm{e}}(0)$ translates to changes of \unit[$\sim$0.22]{GV/m} or \unit[$\sim$0.5]{\%}, much smaller than the variations that are typically observed in experiments~\cite{ebeamsbella,beamsfromdischargecap3}.

The two major components of laser driven plasma wakefield accelerators are the laser and the plasma. Our measurement results highlight that discharge plasma waveguide parameter variations are small compared to typical variations of the pulses produced by LPA laser systems. Thus, increased laser pulse parameter reproducibility is essential to achieve significant improvements in the reproducibility and stability of particle bunches, consistent with the findings of~\cite{repDESY}.

\subsection{Plasma Telescopes}
\label{sec:plasmatelescope}
Waveguides could also be used as 'plasma telescopes': for the right combination of capillary length and matched spot size the pulse exits the plasma after half a spot-size oscillation, changing it from $w_\mathrm{0}$ to $w_\mathrm{m}^2/w_\mathrm{0}$. This feature may be used to increase or decrease the spot size of an high-intensity pulse in a short distance, where conventional optics would be destroyed. 

\begin{figure}[htb!]
    \centering
    \includegraphics[width=0.8\textwidth]{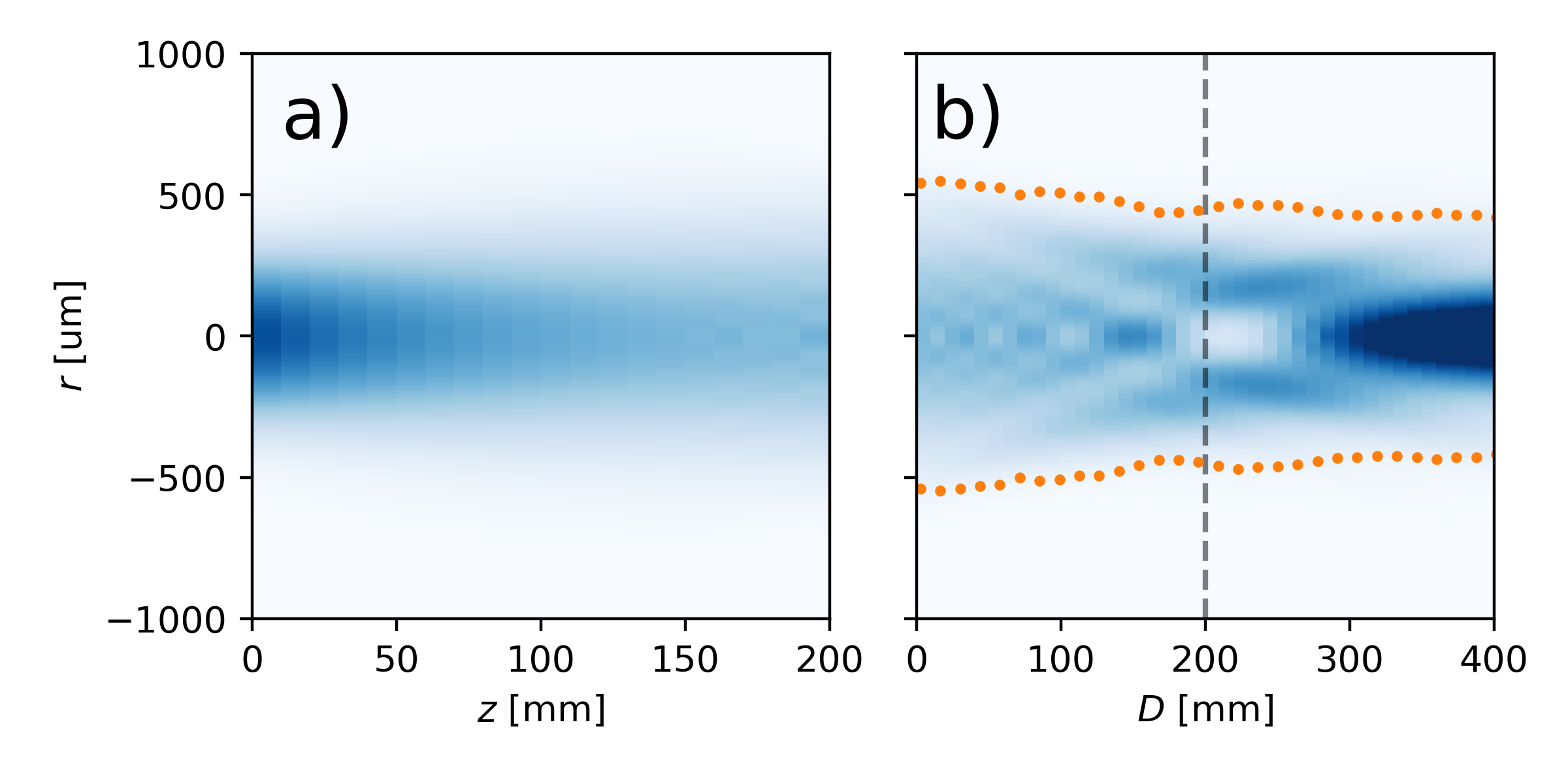}
    \includegraphics[width=0.9\textwidth]{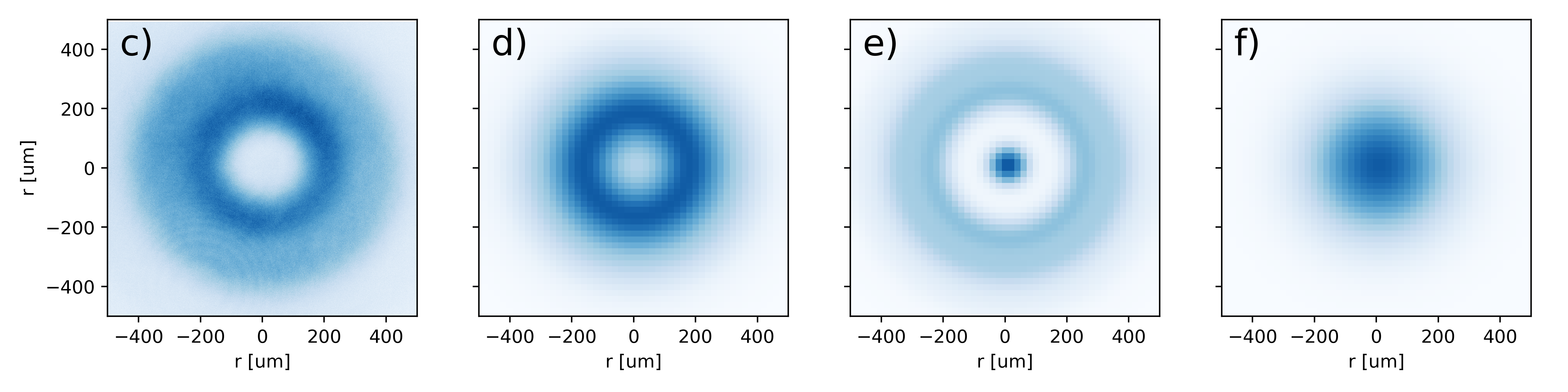}
    \caption{a) Simulation result of the laser pulse intensity evolution along the capillary~\#3 in a plasma telescope configuration, using the channel profile according to Fig.~\ref{fig:hextranschannel} as input. b) simulated pulse intensity evolution  downstream the capillary exit $D$. Orange markers show the second moment of the pulse obtained from experimental measurements. The gray vertical dashed line indicates the location of the measured intensity distributions shown in panels c-f. c) the experimentally measured intensity profile, d) the corresponding simulation result when using the Gerchberg-Sexton algorithm to reconstruct the input pulse modes, e) when using a perfect Gaussian pulse as simulation input and f) a Gaussian pulse that was matched to the channel with \unit[$w_{\mathrm{0}}=245$]{$\mu$m}.}
    \label{fig:resize}
\end{figure}

To understand how the $r^4$ component in Eq.~\ref{eq:channel} affects pulse propagation for waveguides that are used as plasma telescopes, we performed measurements on the \unit[2]{mm} diameter capillary~\#3. We placed the capillary entrance \unit[20]{cm} downstream the pulse focus (\unit[$w_{\mathrm{0}}\sim$105]{$\mu$m}) and used the waveguide (see Fig.~\ref{fig:hextranschannel}) to 'collimate' the diverging pulse. Measurements of the pulse profile downstream the waveguide showed that the pulse is no longer transversely Gaussian (see Fig.~\ref{fig:resize}c). A radially symmetric ring structure was clearly visible. The pulse spot size (calculated from the second moment of the transverse intensity distribution) oscillated downstream the waveguide as shown by orange points in Fig.~\ref{fig:resize}b. 

We simulated pulse propagation with INF\&RNO, using the reconstructed input pulse and the NPINCH channel profile from Fig.~\ref{fig:hextranschannel}. Figure~\ref{fig:resize}a shows the pulse evolution inside the capillary and Fig.~\ref{fig:resize}b downstream the capillary exit. The transverse pulse intensity distribution downstream the capillary exit showed higher order modes, qualitatively similar to the experimental measurements; the pulse peak intensity as well as its spot size oscillated as different modes came in and out of focus. Figures~\ref{fig:resize}c-e compare transverse pulse intensity distributions \unit[20]{cm} downstream the capillary exit. Figure~\ref{fig:resize}c is the experimentally measured distribution; Fig.~\ref{fig:resize}d was obtained from simulations when using a model of the input pulse and Fig.~\ref{fig:resize}e from simulations when using a transversely Gaussian input pulse. Figures~\ref{fig:resize}c-e all show the same 'ring' features, even when the input pulse is a perfect Gaussian (Fig.~\ref{fig:resize}e); the 'ring' features are therefore a result of the pulse experiencing the non-parabolic contributions of Eq.~\ref{eq:channel}. These initial experiments highlighted that R\&D will be required before capillary discharge waveguides can be used as plasma telescopes for applications that require a Gaussian mode.

Figure~\ref{fig:resize}f shows the simulated transverse pulse intensity distribution \unit[20]{cm} downstream the capillary exit for a pulse (\unit[$w_\mathrm{0}=245$]{um}) that was matched to the plasma channel (\unit[$w_\mathrm{m}=245$]{um}, $E_\psi=0.12$). Non-Gaussian features were negligible. This simulation result confirms that when $w_\mathrm{0}\sim w_\mathrm{m}$, the effects of non-parabolic channel contributions on pulse propagation are insignificant. 

\section{Conclusions \& Summary}
We evaluated the reproducibility of plasma channels formed by discharges in gas-filled capillary waveguides and obtained precise measurements of their radial plasma electron density distributions. Variations of the waveguide central axis were below the \unit[1]{$\mu$m} resolution of the measurement; variations of the average on-axis density had an upper limit of \unit[1]{\%}. Matched spot size variations were largest (\unit[$\Delta w_\mathrm{m}=$0.2]{\%}) for the longest capillary (\unit[40]{cm}).  We project that these quantities are not sufficient to cause the large electron beam parameter variations that are typically observed in laser driven plasma wakefield acceleration experiments to date~\cite{ebeamsbella,beamsfromdischargecap3}. Improvements of the quality and reproducibility of particle bunches accelerated in laser driven plasma wakefields therefore require improvements of the laser pulse parameter variations~\cite{repDESY}. 

Measurements of the transverse waveguide profiles revealed non-parabolic contributions to the radial channel profile that were in excellent agreement with magneto-hydro-dynamic simulation results. We demonstrated that non-parabolic effects influenced pulse propagation for non-matched waveguide applications, such as plasma telescopes. However, when the pulse was approximately matched to the waveguide, their effects were negligible. 

\section{Acknowlegements}
This work was supported by the Director, Office of Science, Office of High Energy Physics, of the U.S. Department of Energy under Contract No. DE-AC02-05CH11231, and used the computational facilities at the National Energy Research Scientific Computing Center (NERSC), as well as the project High Field Initiative (No. CZ.02.1.01/0.0/0.0/15\_003/0000449) from the European Regional Development Fund.
The authors gratefully acknowledge the technical support from Zac Eisentraut, Mark Kirkpatrick, Arturo Magana, Greg Mannino, Joe Riley, Tyler Sipla and Nathan Ybarrolaza.

\cleardoublepage


\begin{thebibliography}{99}
\bibitem{uniformitychannel} We monitor longitudinal uniformity by imaging the light emitted by the discharge onto the gated camera shown in Fig.~\ref{fig:setup} (gate time 3 ns).
\end{thebibliography}
\end{document}